\documentclass{appolb}
\usepackage{graphicx}
\usepackage{amssymb}
\usepackage{epsfig}

\begin{document}

\title{Resolved photon and multi-component model \\
       for $\gamma^*$p and $\gamma^* \gamma^*$ total cross section
       \thanks{Presented at Photon2005, Warsaw, August/September 2005}
       }

\author{A. Szczurek$^{1},^{2}$ and T. Pietrycki$^{1}$ \\
{\it $^{1}$ Institute of Nuclear Physics,
         PL-31-342 Cracow, Poland } \\
{\it $^{2}$ University of Rzesz\'ow, 
         PL-35-959 Rzesz\'ow, Poland } \\ }
\maketitle

\begin{abstract}
We generalize our previous model for $\gamma^* p$ scattering
to $\gamma \gamma$ scattering. Performing 
a new simultaneous fit to $\gamma^* p$ and $\gamma \gamma$ total
cross section we find an optimal set of parameters to describe 
both processes. 
We propose new measures of factorization breaking for $\gamma^*
\gamma^*$ collisions and present results for our new model.
\end{abstract}

\section{Introduction}

In the last decade the photon-proton and photon-photon
reactions became a testing ground for different QCD-inspired
models. The dipole model was one of the most popular and successful.
In the simplest version of the model
only quark-antiquark Fock components of the photon are included
in order to describe the total cross sections.
In contrast, the more exclusive processes, like diffraction \cite{GBW_glue},
jet \cite{jets_resolved_photon} or heavy quark \cite{AS02} production, require
inclusion of higher Fock components of the photon.

In Ref. \cite{PS03} we have constructed a simple hybrid model
which includes the resolved photon component in addition to
the quark-antiquark component. With a very small number of parameters
we were able to describe the HERA $\gamma^* p$ total cross section data 
with an accuracy similar to that of very popular dipole models.
In Ref. \cite{PS05} we have generalized our hybrid model also
to photon-photon collisions.
Application to $\gamma \gamma$ processes requires some modifications
of the model.

\section{Formulation of the model}

\begin{figure}[htb] 
\begin{center}
\includegraphics[width=6cm]{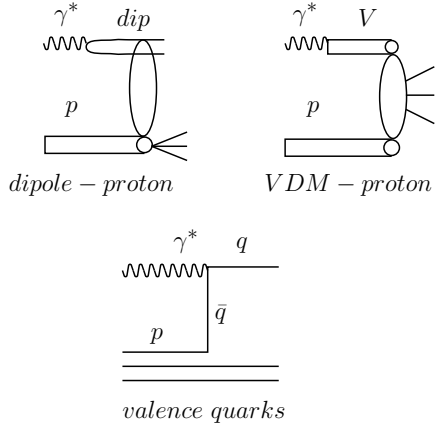}
\caption{\it
The graphical illustration of the multicomponent $\gamma^* $p 
scattering model.
\label{fig:fig1}
}
\end{center}
\end{figure}

In our model the total cross section for $\gamma^* p$ is a sum 
of three components illustrated graphically in Fig.\ref{fig:fig1}.

For the dipole component
\begin{equation}
\sigma^{tot}_{dip}(W,Q^2) = \sum_q \int dz \int d^2\rho \;\sum_{T,L}
\left|\Psi^{T,L}_{\gamma^*\rightarrow q \bar q}(Q,z,\rho) \right|^2
\cdot \sigma_{(q \bar q)N}(x,\rho)  
\end{equation}
and for the vector meson component
\begin{equation}
\sigma^{tot}_{VDM}(W,Q^2) = \sum_V \frac{4\pi}{\gamma^2_V}
\frac{M^4_V \sigma^{VN}_{tot}(W)}{(Q^2 + M_V^2)^2} \cdot (1 - x) . 
\end{equation}
The last component in Fig.\ref{fig:fig1} becomes important only at large $x$, i.e. small W.

We take the simplest diagonal version of VDM with $\rho$, $\omega$ and
$\phi$ mesons included. The vector-meson-nucleon cross section is
approximated by pion(kaon)-proton cross section. 
A simple Regge parametrization by Donnachie and Landshoff \cite{DL92} is used
to parametrize the pion(kaon)-proton total cross section.
We take $\gamma$'s calculated from the leptonic decays
of vector mesons, including finite width corrections.

\begin{figure}[htb] 
\begin{center}
\includegraphics[width=6cm]{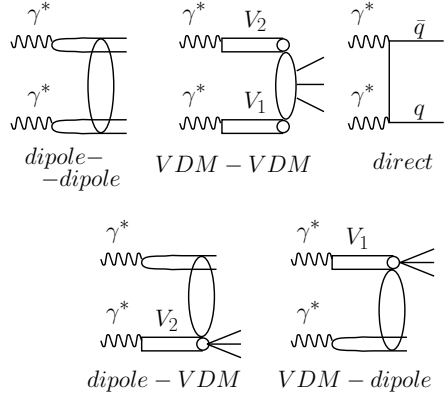}
\vspace{-0.1cm}
\caption{\it
The graphical illustration of the multicomponent $\gamma^* \gamma^*$ 
scattering model.
\label{fig:fig2}
}
\end{center}
\end{figure}

In the same spirit, the total cross section for $\gamma^* \gamma^*$ scattering
can be written as a sum of five terms shown in Fig.\ref{fig:fig2}.

The formulae for the direct term can be found in Ref.\cite{Budnev}.

If both photons fluctuate into perturbative quark-antiquark pairs,
the interaction is due to gluonic exchanges between quarks and antiquarks
represented in Fig.\ref{fig:fig2} by the blob.
Formally this component can be written in terms of the photon
perturbative "wave functions" and the cross section for the interaction
of both dipoles
\begin{eqnarray}
\sigma^{tot}_{dip-dip}(W,Q^2_1,Q^2_2)&=& 
\sum^{N_f}_{a,b=1}
\int_0^1dz_1\int d^2\rho_1|\Psi^a_T(z_1,\rho_1)|^2 \nonumber \\
&\cdot&\int_0^1dz_2\int d^2\rho_2
|\Psi^b_T(z_2,\rho_2)|^2 
\sigma^{a,b}_{dd}({\bar x}_{ab},\rho_1,\rho_2). \\ \nonumber   
\label{tot_dipdip}
\end{eqnarray}
In paper \cite{TKM02} a phenomenological
parametrization for the azimuthal-angle averaged dipole-dipole
cross section has been proposed:
\begin{equation}
\sigma_{dd}^{a,b}(x_{ab},\rho_1,\rho_2) =
\sigma_0^{a,b}
\left [
 1 - \exp\left(- \frac{\rho_{eff}^2}{4 R_0^2(x_{ab})}  \right)
\right ] 
\cdot S_{thresh}(x_{ab}) \; .
\label{saturation_parametrization}
\end{equation}
Our formula for $x_{ab}$ is different from the one used in 
Ref.\cite{TKM02}. As discussed in Ref.\cite{AS02} our formula 
provides correct behaviour at threshold energies.
Different prescriptions for $\rho_{eff}$ have been considered
in Ref.\cite{TKM02}, with $\rho_{eff}^2 = \frac{\rho_1^2 \rho_2^2}{\rho_1^2 +
  \rho_2^2}$ being probably the best choice \cite{TKM02}.

Following our philosophy of explicitly including the
nonperturbative resolved photon, in photon-photon collisions completely 
new terms must be included (the last two diagrams in
Fig.\ref{fig:fig2}).
If one of the photons fluctuates into a quark-antiquark dipole and
the second photon fluctuates into a vector meson, or vice versa
\begin{eqnarray}
\sigma^{tot}_{SR1}(W,Q^2_1,Q^2_2) &=& 
\int d^2\rho_2\int dz_2 \sum_{V_1}\frac{4\pi}{f^2_{V_1}}
   \left(\frac{m^2_{V_1}}{m^2_{V_1}+Q^2_1}\right)^2 \cdot \nonumber \\ 
&\cdot& \left|\Psi(\rho_2,z_2,Q^2_2)\right|^2 
\sigma^{tot}_{V_1d}(W,Q^2_2)  \; , 
\end{eqnarray} 
\begin{eqnarray}
\sigma^{tot}_{SR2}(W,Q^2_1,Q^2_2) &=& 
\int d^2\rho_1\int dz_1 \sum_{V_2}\frac{4\pi}{f^2_{V_2}}
   \left(\frac{m^2_{V_2}}{m^2_{V_2}+Q^2_2}\right)^2 \cdot \nonumber \\ 
&\cdot& \left|\Psi(\rho_1,z_1,Q^2_1)\right|^2 
\sigma^{tot}_{V_2d}(W,Q^2_1) \; .  
\end{eqnarray} 
In the formulae above:
\begin{equation}
\sigma_{V_i d}^{tot}(W,Q^2) = \sigma_0 \left(1-\mathrm{exp}
\left(-\frac{\rho_i^2}{4R_0^2(x_g)}\right) \right)\cdot S_{thresh} \; .
\end{equation}
In the present calculation we take $m_f = m_0$
for $u / {\bar u}$ and $d / {\bar d}$ (anti)quarks and $m_f = m_0 +$ 0.15
GeV for $s/{\bar s}$ (anti)quarks.

If each of the photons fluctuates into a vector meson the
corresponding component is called double resolved. 
The corresponding cross section reads:
\begin{eqnarray}
\sigma^{tot}_{DR}(W,Q^2_1,Q^2_2) &=&
\sum_{V_1 V_2} \frac{4\pi}{f^2_{V_1}}
  \left(\frac{m^2_{V_1}}{m^2_{V_1}+Q^2_1}\right)^2 \cdot \nonumber \\
&\cdot& \frac{4\pi}{f^2_{V_2}}
  \left(\frac{m^2_{V_2}}{m^2_{V_2}+Q^2_2}\right)^2
  \sigma^{tot}_{V_1 V_2}(W) \; .  
\end{eqnarray} 

The total cross section for $V_1$-$V_2$ scattering
must be modeled. In the following we assume
Regge factorization and use a simple parametrization which
fits the world experimental data for hadron-hadron
total cross sections \cite{DL92}. More details can be found in Ref.\cite{PS05}.

\section{Results}

\begin{figure}[htb] 
\begin{center}
\includegraphics[width=6cm]{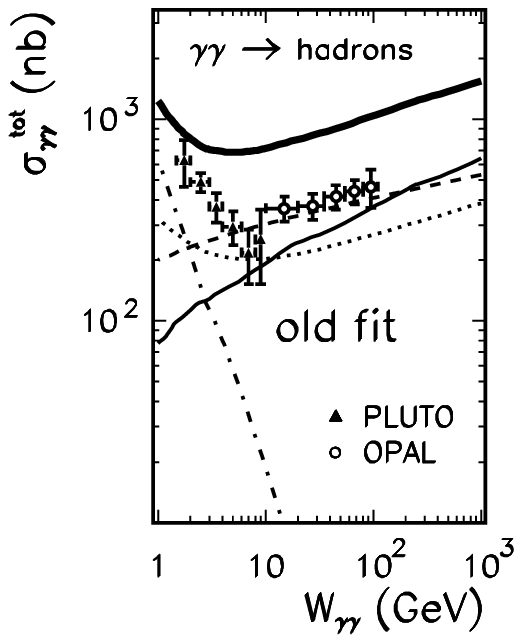}
\includegraphics[width=6cm]{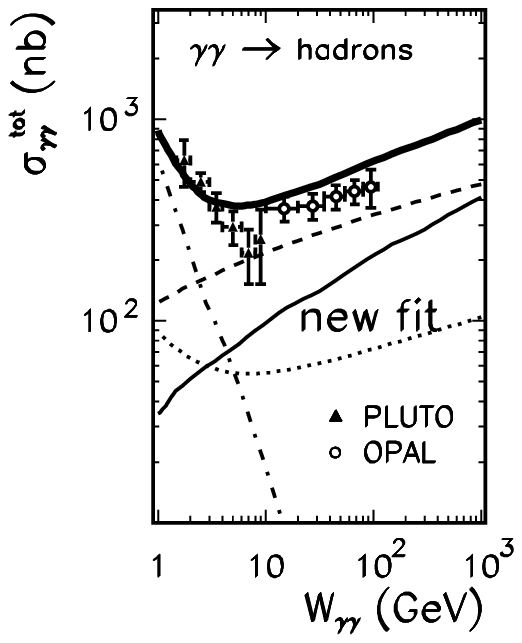}
\vspace{-0.5cm}
\caption{\it  
The total $\gamma \gamma$ cross section as a function 
of photon-photon energy with parameters from Ref.\cite{PS03} (panel a)
and with the new set of parameters.
The experimental data are from \cite{PLUTO,OPAL}.
\label{fig:gg_tot}
}
\end{center}
\end{figure}

In Ref.\cite{PS03} we have adjusted the parameters of our model
to $\gamma^* p$ collisions. Let us try to use these parameters
to describe $\gamma \gamma$ total cross section.
In Fig.\ref{fig:gg_tot} we present the total cross section
as a function of center-of-mass energy.
The sum of all components of Fig.\ref{fig:fig2} (thick-solid line) exceeds
the experimental data by a factor of two or even more. The individual
components are shown explicitly as well. The direct component
(dash-dotted line) dominates at low energies only. At high energies 
the dipole-dipole (thin-solid line), single-resolved (dashed line) and 
double-resolved (dotted line) components are of comparable size.
The overestimation of the experimental data suggests a double-counting.

In Ref.\cite{PS03} it was assumed that the coupling constants
responsible for the transition of photons into vector mesons
are the same as those obtained from the leptonic decays of vector
mesons, i.e. the on-shell approximation was used.
In our case we need the corresponding coupling constants rather at
$Q^2$ = 0 and not on the meson mass shell ($Q^2 = m_V^2$).
We replace $\frac{4\pi}{f^2_{V_i}}\rightarrow 
\frac{4\pi}{f^2_{V_i}}F_{off}(Q^2,m^2_{V_i}) $ and 
extrapolate from meson mass shell to $Q^2$ = 0 by means of
\begin{equation}
F_{off}(Q^2,m^2_{V_i}) = \mathrm{exp}\left(-\frac{(Q^2 + m^2_{V_i})}
{2\Lambda_E^2} \right)
\; . 
\label{F_off}
\end{equation}
The parameter $\Lambda_E$ is a new nonperturbative parameter 
of our new model.
Secondly, the ``photon-wave functions'' commonly used in the 
literature allow for large quark-antiquark dipoles. 
This is a nonperturbative region
where the pQCD is not expected to work. Furthermore this is
a region which is taken into account in the resolved
photon components as explicit vector mesons.
Therefore we propose the following
modification of the ``perturbative'' photon wave function:
\begin{equation}
\left|\Psi(\rho,z,Q^2) \right|^2 \rightarrow \left|\Psi(\rho,z,Q^2) \right|^2
\mathrm{exp}\left(- \frac{\rho}{\rho_0} \right) \: .
\label{dipole_size_modification}
\end{equation}

The parameters $\Lambda_E$ and $\rho_0$ were obtained by fitting our
modified model formula to the experimental data.
The $\gamma \gamma$ data is not sufficient for this purpose
as different combinations of the two parameters lead
to equally good description. Therefore we were forced to perform a
new fit of the model parameters to both $\gamma^* p$
and $\gamma \gamma$ scattering.

\begin{figure}[htb] 
\vspace{-0.5cm}
\begin{center}
\includegraphics[width=5.5cm]{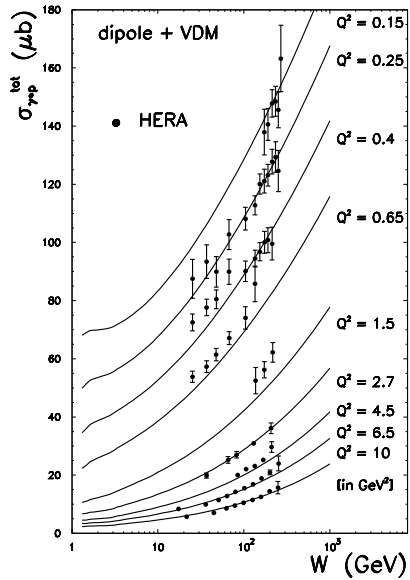}
\caption{\it
The total $\gamma^* p$ cross section as a function 
of photon-proton energy. The experimental HERA data 
are from \cite{HERA_data}.
\label{fig:hera}
}
\end{center}
\end{figure}

In Fig.\ref{fig:gg_tot} we show the resulting $\sigma_{tot}^{\gamma \gamma}$
together with the experimental data of the PLUTO (solid triangles) and
OPAL (open circles)
collaborations. We also show the individual contributions of different
processes from Fig.\ref{fig:fig2}.
The relative size of the contributions has changed when compared
to the old set of parameters. Now the sum of the 
single resolved components, included here for the first time,
dominates in the broad range of center-of-mass energies.
The double resolved component is now much weaker and constitutes 10-15 \%
of the total cross section only.
In Fig.\ref{fig:hera} we show the analogous description
of the $\gamma^* p$ data. The agreement with the HERA data is similar 
as in our previous paper \cite{PS03}.

In our fit we have included $\gamma^* p$
and $\gamma \gamma$ experimental data. In Ref.\cite{PS05}
we have compared the predictions of our
model for total cross sections for one virtual -- one real photon
with existing experimental data.
 
In data processing, in particular in extrapolations to small 
photon virtualities one often assumes the following relation
\begin{equation}
\sigma_{\gamma^* \gamma^*}^{tot}(W,Q_1^2,Q_2^2)
= \Omega(Q_1^2) \cdot \Omega(Q_2^2) \cdot \sigma(W)
\label{factorization}
\end{equation}
known as factorization. 
We have considered two quantities which measure factorization breaking.
They read:
\begin{eqnarray}
f_{fb}^{(1)}(W,Q_1^2,Q_2^2) &&\equiv
\frac{\sigma_{\gamma^* \gamma^*}(W,Q_1^2,0) \;
      \sigma_{\gamma^* \gamma^*}(W,0,Q_2^2)}
     {\sigma_{\gamma^* \gamma^*}(W,Q_1^2,Q_2^2) \;
      \sigma_{\gamma^* \gamma^*}(W,0,0)} \; , \nonumber \\
f_{fb}^{(2)}(W,Q_1^2,Q_2^2) &&\equiv
\frac{\sigma_{\gamma^* \gamma^*}(W,Q_1^2,Q_1^2) \;
      \sigma_{\gamma^* \gamma^*}(W,Q_2^2,Q_2^2)}
     {\sigma_{\gamma^* \gamma^*}(W,Q_1^2,Q_2^2) \;
      \sigma_{\gamma^* \gamma^*}(W,Q_2^2,Q_1^2)}  \; .
\label{f_fb}
\end{eqnarray}
For the factorized Ansatz (\ref{factorization})
$f_{fb}^{(1,2)}(W,Q_1^2,Q_2^2)$ = 1 for any $Q_1^2$ and $Q_2^2$.

\begin{figure}[htb] 
\vspace{-0.5cm}
\begin{center}
    \includegraphics[width=6cm]{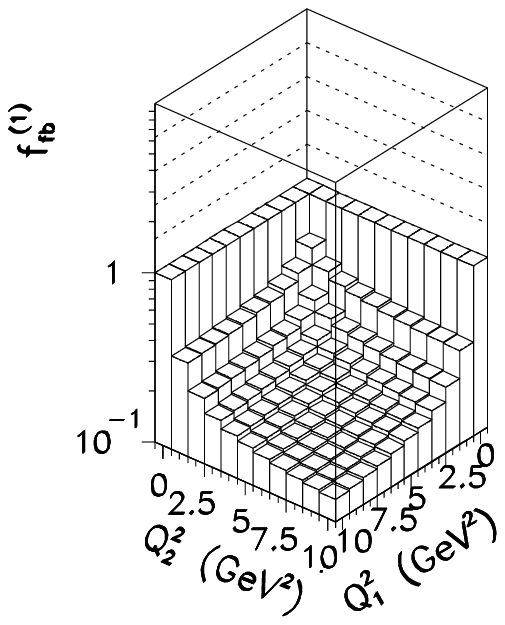}
    \includegraphics[width=6cm]{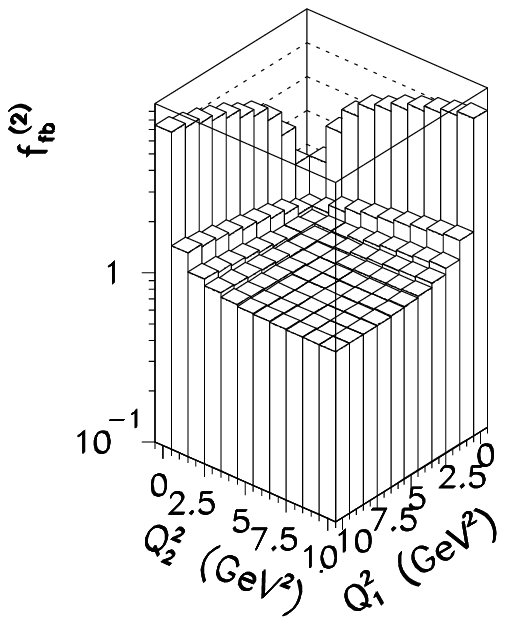}
\caption{\it
The maps of the factorization-breaking functions $f_{fb}^{(1,2)}$ 
as a function of both photon virtualities $Q_1^2$ and $Q_2^2$ for
$W = 100$ GeV.
\label{fig:fb}
}
\end{center}
\end{figure}

The factorization-breaking functions $f_{fb}^{(1,2)}$ are shown in
Fig.\ref{fig:fb}
as a function of both photon virtualities $Q_1^2$ and $Q_2^2$ for 
W = 100 GeV.
According to the definitions at $Q_1^2 = 0$ or $Q_2^2 = 0$
$f_{fb}^{(1)}$ = 1 and $f_{fb}^{(2)}$ = 1 when $Q_1^2 = Q_2^2$.

\section{Conclusions}

We have generalized our previous model for $\gamma^* p$ total cross
section to the case of $\gamma \gamma$ scattering. In the last case a
few new components appear.

The naive generalization of our former model for $\gamma^* p$ total cross
section leads to a serious overestimation of the $\gamma \gamma$ total
cross sections. A priori, this fact can be due either to a nonoptimal
set of model parameters found in our previous study, double counting, or
due to some model simplifications like off-shell effects. 
We have suggested to include such an effect 
by introducing new form factors. 
When including the quark-antiquark 
continuum one usually takes into account the perturbative quark-antiquark 
"photon wave function". This is justified and reasonable for small
size dipoles only. In order to avoid double counting the large-size
dipoles have been eliminated using a simple exponential function 
in transverse dipole size. We have performed 
a new fit of our generalized-model parameters to the $\gamma^*$p and 
$\gamma \gamma$ total cross sections.

When trying to extrapolate the experimental cross sections for
the $\gamma^* \gamma^*$ scattering to real photons one often 
assumes factorization.
We have quantified the effects of factorization breaking in our 
model with parameters fixed to describe the $\gamma^* p$ and 
$\gamma \gamma$ data. 
We have proposed two new functions which can
be used as a measure of the factorization breaking.



\end{document}